# LIGHT WILL BE THROWN: THE EMERGING SCIENCE OF CULTURAL EVOLUTION

Chris Buskes

Department of Philosophy, Radboud University, Nijmegen, The Netherlands
c.buskes@ftr.ru.nl

## ABSTRACT

*Culture evolves, not just in the trivial sense that cultures change over time, but also in the strong sense that such change is governed by Darwinian principles. Both biological and cultural evolution are essentially cumulative selection processes in which information – whether genetic or cultural – is sieved, retained and then transmitted to the next generation. In both domains such a process will result in recognizable lineages and tree-like phylogenies so characteristic of Darwinian evolution. Because a principle of inheritance (i.e., faithful replication of information) holds in both domains, we may trace back particular transmission histories and identify the forces that influenced them. The idea that culture evolves is quite old, but only in recent years there has been a serious effort to turn this idea into science. This article offers a concise analysis of how a rudimentary idea gradually evolved into a thriving research program.*

## 1. INTRODUCTION

On the very last pages of *The Origin of Species*, seemingly written down as a brief aside or afterthought, Darwin (1859) offers a glimpse of a future science that one day will further illuminate our origin and our place in the world. Despite the fact that there are big ideas at stake, the remarks are made casually in but a few short staccato sentences. Darwin famously writes: "In the distant future I see open fields for far more important researches. Psychology will be based on a new foundation... Light will be thrown on the origin of man and his history." Apparently the author of these words envisaged a future in which not only biological sciences like zoology, embryology and paleontology but also the social sciences, the sciences of mind and behavior, and perhaps even some of the humanities would be unified by a new theoretical foundation. What will bind the various scientific disciplines together into a bigger, comprehensive picture is the principle of evolution. A couple of decades after the publication of the *Origin*, Darwin (1871; 1872) himself took up this audacious task by writing two comprehensive volumes on human origins (*The Descent of Man*) and biological psychology (*The Expression of the Emotions in Man and Animals*) respectively. Yet after these first promising endeavors, the project eventually stalled, not to be picked up again by others well until after the Second World War.

As long as there has been a viable theory of evolution, people have been eager to draw meaningful parallels and analogies between biological evolution on the one hand and the workings of human society and culture on the other. Initially these first attempts to "biologize" the sphere of human action were rather primitive and crude, mainly because there were still many misunderstandings about the phenomenon of evolution. Thus during the second half of the 19$^{th}$ century Herbert Spencer, the founder of Social Darwinism, argued that the theory of evolution could also be applied to human society. Spencer believed that evolution involves fierce competition between individuals – he famously coined the immortal phrase "survival of the fittest" – and that the government should not intervene in this natural elimination process. A hard but fair struggle for existence will lead to progress because in the end the strong and intelligent individuals will persevere whereas the weak and dumb will deservedly perish. Some decades later, in the first half of the 20$^{th}$ century, things got even worse. With the advent of eugenics and obscure doctrines about alleged superior and inferior races, Darwin's theory of evolution became entangled and tainted with Nazi ideology. As we all know, this sinister type of pseudoscience eventually lead to suppression, persecution and genocide. Not surprisingly then that after the Second World War it became a real taboo to "Darwinize" or "biologize" human behavior, society and culture in any way. Although, admittedly, there were a few exceptions. Thus in the 1960's the Austrian ethologist Konrad Lorenz (1966) tried to explain several aspects of human behavior – most notably aggression against members of the same species – in terms of biological instincts, while during the same period the American psychologist Donald Campbell (1960) was proposing Darwinian ideas of cultural and scientific evolution (he also coined the term "evolutionary epistemology"). Yet on the whole there was not much readiness to take these biological and evolutionary approaches very seriously because, at that time and especially in the social sciences, the influence of culture on the human condition was deemed much more important that the influence of nature.

It was not until the mid-1970's that things cleared up a bit. In 1975 the renowned Harvard biologist Edward O. Wilson published his book *Sociobiology*, a seminal work on the biological and evolutionary basis of animal and human social behavior. Initially Wilson's ideas were heavily criticized and sneered at, but within a decade or two Wilson's sociobiology became generally accepted as good science and the proper way to study animal and human communities. In Wilson's wake, a new research program gradually emerged in which animal and human communities were studied from an evolutionary perspective. Later, with the birth of evolutionary psychology, the study of the human (and animal) mind became incorporated into this program as well (Barkow, Cosmides and Tooby 1992). And since the turn of the century we have witnessed the rapid rise of yet another related and promising branch of research: the science of cultural evolution, the main topic of this article. The science of cultural evolution is a relatively new, highly collaborative and multidisciplinary enterprise which harbors evolutionary biologists, archaeologists, (paleo)anthropologists, psychologists, sociologists, linguists and philosophers among others (see: Levinson and Jaisson (Eds.) 2006; Whiten et al. (Eds.) 2012; Richerson and Christiansen (Eds.) 2013). This disciplinary diversity is no coincidence since many participants believe that the science of cultural evolution will unify the sciences in unprecedented ways (Mesoudi, Whiten and Laland 2006; Buskes 2015). Indeed, it may well be that we are at the brink of a new era, an era in which the Darwinian paradigm perhaps will reach its full potential.

## 2. CO-EVOLUTION AND NICHE CONSTRUCTION

We have seen that after a few false starts and a long delay, the topic of evolution and human behavior finally became accepted as proper science in the 1970's, thanks to the pioneering work of Ed Wilson and others. Wilson's seminal work *Sociobiology* paved the way for a new research program in which our species was treated as just another animal, and in which human social behavior was studied within the framework of biology and evolution. Consider for instance the example of cooperative and altruistic behavior that we witness in many species, most notably in social insects like ants and bees, but also in humans. The science of sociobiology, in brief, postulates two general mechanisms for explaining such sustained behavior, to wit: kin selection and reciprocity. Kin selection means that altruistic behavior might persist if the behavior is directed towards close kin, and reciprocity means that altruistic behavior can be a good and stable strategy if it is based on the principle "if you scratch my back, I'll scratch yours". This was all sound biology first introduced by Bill Hamilton (1964) and Robert Trivers (1971), respectively. The initial controversy lay somewhere else. What had enraged early critics was, first, Wilson's insistence that his sociobiological model was not only applicable to animals but also to humans, and second, that he apparently defended a bold gene-centered view of evolution. The first critique was easily countered because Darwin himself already convincingly showed that there is no fundamental "gap" between animals and humans. Despite our human ingenuity and cultural splendor, we are still a species of primate and thus closely related to monkeys and great apes.

Yet the second critique was not so easily dismissed. Apparently Wilson argued that, ultimately, everything in evolution revolves around the propagation of genes, as was demonstrated some years later when he (in)famously wrote that "genes hold culture on a leash" (Wilson 1978: 167). Sociobiologists like Wilson try to understand human behavior, the human mind and human culture chiefly – some would say solely – in terms of our biological constitution. Elsewhere I have called this idea the "Nature-First approach" (Buskes 2013: 677). This view proclaims that human culture is embedded within the wider and much older framework of human nature. Wilson insists that our so-called "epigenetic rules", i.e., our evolved mental capacities and psychological dispositions, will ultimately bias and constrain the ways in which our culture can evolve. So cultures which disobey the imperatives of our biology (our genes) will either be pulled back by the leash or go extinct.

At first sight the Nature-First approach has a certain appeal because no sensible person would deny that we indeed *are* biological creatures, and that an understanding of genetic and epigenetic constraints on human behavior and human culture is therefore simply indispensible. The flaw of the Nature-First approach is that it underestimated the major role and increasing influence that culture has played in our evolutionary history. For cultural evolution is not merely a process that is embedded within – and thus to a certain extent subordinate to – biological evolution, but is rather intertwined with it. Cultural changes could bring about alterations to the environment which in turn affect both how genes act in development and what selection pressures act on genes. The classic example is adult lactose tolerance among human populations (Durham 1991). Lactose tolerance in certain, mostly European, populations has arisen as a consequence of a cultural innovation our ancestors carried through some 8,000 years ago: dairy farming. From then on adult individuals with the ability to digest milk (and other dairy products) had an advantage over individual who could not, hence a new digestive enzyme got selected. Thus the evolution of lactose tolerance is a fine example of how culture can have a causal impact on our genes. Other examples of how culture or cultural innovations have fed back on our gene pool are our ancestor's preference for eating meat (Aiello and Wheeler 1995), the control of fire (Berna et al. 2012) and the subsequent invention of cooking (Wrangham et al. 1999). It is most likely that without these changing preferences and innovations the spectacular increase of the hominid brain would not have been possible. Because carnivores do not need a long digestive tract, especially when the meat is first cooked, our ancestors could "swap" their excess gut tissue for more tissue of the brain.

More recent research shows that culture must have had, and still has, a huge influence on our genes. More specifically, cultural processes may affect the rate of change of allele frequencies, sometimes speeding it up and sometimes slowing it down, thus gradually altering the make-up of the human genome. The list of genes that have been subject to, and thus have been altered by, cultural selection pressures is still growing. It includes genes that predispose individuals to learn (and to learn from others), genes that affect personality, intelligence, and lateralization of hand preference, genes that affect dietary preferences and alcohol metabolism, genes that affect hair and eye color, and genes (e.g., the FOXP2 gene) that have facilitated the acquisition of language (Laland, Odling-Smee and Myles 2010).

In short, many cultural practices and innovations have changed our genetic makeup and hence our physiology, and these examples convincingly show that cultural evolution is not at all subordinate to biological-genetic evolution because culture can alter the path of human evolution. This intricate process is also known as "niche construction", a process of environmental modification through which organisms alter the selection pressures acting on themselves and other organisms (Laland, Odling-Smee and Feldman 2000). By setting up and advancing their culture, our ancestors created a new "niche" in which it paid off to have big, "expensive" brains. Contrary to animal niche construction (the classic example is the beaver dam), human niche construction is characterized by social learning and cultural inherited practices which have extensively modified the biological selection pressures acting on our species.

In short, because culture can change the path of biological evolution, the Nature-First approach had to be adjusted. Thus, in their effort to revalue the role of culture, Cavalli-Sforza and Feldman (1981) and Lumsden and Wilson (1981) proposed a model of "gene-culture co-evolution" in which genes and culture are inextricably connected. So instead of being subordinate to human nature, culture was now rightly deemed an evolutionary force in its own right. A further big improvement was made by Robert Boyd and Peter Richerson (1985), two pioneering researchers in the science of cultural evolution, when they introduced their model of "dual inheritance", i.e., the idea that culture not only feeds back on our genes, but that culture has its own transmission channel or "inheritance system" as well. Dual inheritance means that, in principle, there are two ways in which information can be transmitted to the next generation, either by genetic replication (i.e., sexual reproduction) or by imitation and language. As we shall see further on, both kinds of transmission are essentially Darwinian cumulative selection processes in which particular variants (whether genetic or cultural) are selectively retained, thus resulting in a gradual accumulation of information. A novel cultural inheritance system arose next to the older genetic one because cultural evolution was able to cope much faster with environmental changes than biological-genetic evolution. I will delve deeper into this matter in the next section.

Some authors have argued that the new facts and insights about gene-culture co-evolution and cultural inheritance require an "extended evolutionary synthesis" that would update and "modernize" the modern synthesis of evolution (Danchin et al. 2011). The term "modern synthesis" refers to the unification of the biological sciences during the 1920's and 1930's, most notably the merging of Darwin's theory of natural selection and Mendel's theory of inheritance, into a single biological framework which became known as "Neo-Darwinism". Several decades later, in the 1970's, Wilson gave his book *Sociobiology* the telling subtitle "the new synthesis", now urging for a further unification of the biological sciences and the social sciences. With the advent of the science of cultural evolution and cultural inheritance such a new synthesis – an extended theory of evolution – is perhaps within reach. Thus the evolutionary anthropologist Alex Mesoudi (2011) argues that the theory of cultural evolution will synthesize the social sciences in the same way as the theory of biological evolution has synthesized the biological sciences in the 1920's and 1930's (see also Mesoudi et al. 2006). Mesoudi argues that up to now the social sciences were hampered by the fact that they apply different methods and techniques, hold mutually incompatible theoretical assumptions, and study different subjects. The science of cultural evolution will surmount these traditional disciplinary divisions by providing a common theoretical framework which in turn will spark off the long-awaited unification. This may be a bold claim and a big promise, but perhaps the same would have been said to somebody at the beginning of the 20$^{th}$ century who proclaimed an imminent unification of the biological sciences.

## 3. A NEW TRANSMISSION CHANNEL

The *El Castillo* cave in the Cantabria region in Northern Spain is famous for its many ochre and scarlet hand stencils. The cave walls are covered with dozens of handprints, many of which looking as fresh and colorful as if they were put there yesterday. Yet these prints or stencils, as was found out by uranium-thorium dating, are in fact more than 40,000 years old, several thousand years older than the equally stunning paintings in the Chauvet cave in Southern France, thus making the prints the oldest known examples of Paleolithic cave art (and symbolic thought) in Europe, and perhaps even in the world (Pike et al. 2012). Until recently it was generally assumed that the stencils were made by men, presumably hunters or shamans. But recent research based on the sexual dimorphism with respect to finger length (women's ring and index fingers are about the same length, whereas men's ring fingers tend to be longer than their index fingers) suggests that about 75 percent of the handprints were actually made by women (Snow 2013). Even more intriguing, the age of the figures makes it theoretically possible that the stencils were in fact not made by our Pleistocene ancestors but by our more distant cousins the Neanderthals, because 40,000 years ago the Iberian peninsula was one of their last remaining strongholds in Europe. So assuming that the sexual dimorphism with respect to finger length in Neanderthals was roughly similar to that of modern humans, it is conceivable that most of the red handprints were actually made by Neanderthal women. Of course, this hypothesis is hard to prove and we probably will never know for certain who was responsible for these ancient artistic expressions, but what we *do* know is that the stencils on the cave wall are a clear sign that their makers, whether Neanderthal or human, possessed culture, language and symbolic thought. Symbolic thought is the ability to use symbols, signs or words to represent things. Such an ability requires a considerable mental and social feat because one needs to learn to grasp the (hidden) meaning of a sign or symbol. Thus, to their makers, a crude two-dimensional charcoal sketch on a cave wall may represent real (prey) animals in flesh and blood on the plain. Hand stencils are perhaps more abstract but equally powerful symbols which undoubtedly must have had some (symbolic) meaning to their makers. For the role of symbolic thought in human evolution, see Tattersall (2008). For a recent view on supposed Neanderthal mental, cognitive and linguistic capacities, see Dediu and Levinson (2013).

So the first crude and rudimentary examples of art and symbolic thought can be traced back some 40,000 years. It coincides with what archaeologists and paleoanthropologists often refer to as the "cultural big bang", i.e., the "sudden" appearance of cave paintings, ornaments, figurines and other delicate artifacts in Europe during the Pleistocene. But of course the origins of human – or hominid – culture as such stretch much further back than that. In fact, the evidence in the form of primitive stone tools found in East-Africa – and attributed to *Homo habilis*, or "handy man", the first hominid species belonging to the genus Homo – prove that hominid culture is at least 2,5 million years old. This earliest stone tool industry is called "Oldowan", after the Olduvai Gorge, a famous archaeological site in Tanzania. The Oldowan industry is the earliest tangible evidence of the existence of culture as such, i.e., proof that our ancestors were able to store and accumulate information and transmit it to a next generation. After all, the fabrication of stone tools and artifacts, however primitive, requires skills and knowledge which are generally beyond the level that could be achieved, from scratch, by a single individual or during a single generation. Culture, or at least an evolving culture, requires the *accumulation* of information which causes knowledge and skills to gradually build up (Boyd and Richerson 1995; Tennie et al. 2009). In this way culture opened a new channel – next to the much older genetic transmission channel – through which vital information could be passed on to the next generation. Because that is essentially what "culture" is: information that is transmitted to the next generation by non-genetic means. Furthermore, the emergence of culture made evolution much more complex and dynamic because it now involved the intricate interplay of two replication chains, one genetic and one cultural. And above all, culture made some species to evolve much faster than otherwise would be possible.

To be sure, many nonhuman species, even fish and insects, show cultural traditions in the form of group-specific socially-learned behaviors (Laland and Hoppitt 2003; Kempe et al. 2014). In its most basic form "culture" simply requires that the one individual's behavior (or certain aspects of that behavior) is being imitated by other individuals and then non-genetically transmitted to other members of the group. In this way, a particular (non-congenital) behavioral pattern could then "survive" in the sense that it manages to persist in future generations through social transmission. Social learning is widespread among animal species because it circumvents the often costly, and sometimes deadly, strategy of individual trial-and-error learning (Boyd and Richerson 1985). What is more, the ability to learn from others may well be a necessary condition for the emergence of cultural evolution (Laland 2004; Boyd et al. 2011). In more sophisticated versions "culture" probably requires full-blown language, in the form of symbols or spoken and written word, by which means information can efficiently be stored, accumulated and transmitted. In both scenarios, though, culture has the same basic effect: it allows a species to evolve and to adapt to changing environments much faster than otherwise would be possible. And that brings us right to the question which urges itself upon us: why, in the history of life on our planet, has culture emerged in the first place? What is the apparent advantage of culture?

Now as we have just seen, the answer to the question why culture evolved seems pretty straightforward. Culture evolved simply because it is advantageous in the sense that it makes a species' behavioral repertoire much more flexible. Such adaptive plasticity allows the individuals belonging to that species to survive in a vast range of different habitats. So culture can do things that genes cannot, partly because cultural evolution is much faster than genetic evolution. Take our own species, *Homo sapiens*. Thanks to the advance of culture our prehistoric ancestors were able to control fire and to make tools, weapons, clothes and shelter among other things. These technical innovations in turn allowed our ancestors, within a mere 100,000 years, to conquer and inhabit the far corners of our planet, from the arid deserts and frozen tundras to the most remote islands of the Pacific, and that is quite remarkable for a primate whose body and mind was originally designed to survive on the African savannah. But here's the catch. If culture is as biologically advantageous as it seems to be, why has it not evolved in many other species? In fact, why is *H. sapiens* apparently the only extant species in which culture really evolves? There are more riddles here then one would suspect. Or as Richerson and Boyd (2005: 126) have aptly put it: "The existence of human culture is a deep evolutionary mystery, on a par with the origins of life itself."

## 4. THE MYSTERY OF HUMAN CULTURE

According to Richerson and Boyd, culture is common, but cultural evolution is rare. Although several animal species have managed to acquire culture and sometimes even to generate elaborate cultural traditions, we humans are probably the only example of a species with *cumulative* culture. In our species culture genuinely evolves. The distinction between cumulative and non-cumulative culture is crucially important because, as Mesoudi (2011) points out, only the former constitutes the gradual evolutionary change that Darwin named "descent with modification". That is, only cumulative evolution will result in different cultural lineages and branching family trees. Well-known examples of non-cumulative culture are birdsong (which for the most part has to be learned through imitation), and nut-cracking and termite-fishing among wild monkeys and apes. With regard to birdsong it has even been shown that "dialects" can emerge and persist among local populations, in the same way as a particular human language may contain different dialects spoken by different (sub)groups within the language community. Nonhuman culture, like birdsong, termite-fishing and nut-cracking, is mainly acquired through social learning. Animals with culture are able to exchange and transmit non-genetic information by carefully watching (or listening to) others individuals and then selectively copying their behavior. By contrast, someone who relies on individual learning may acquire knowledge during his or her lifetime while not sharing it with others. So even though such an individual may gather precious knowledge about lots of things, when that individual dies its knowledge is forever lost. Yet although social learning probably is a necessary condition for culture to become cumulative, it is not a sufficient condition because birds, monkeys and chimps *do* engage in social learning but their cultures are nevertheless non-cumulative. Apparently some ingredient has to be added for cultural evolution to evolve.

One would expect that creative invention would do the trick. If you combine social learning *and* creative invention, culture surely would become cumulative – or would it? Alas it probably would not because, like social learning, creative invention may be necessary but not sufficient for culture to become cumulative. For social learning and creative invention to become really effective, and to gradually build up, information has to be faithfully transmitted, i.e., there has to be a process of accurate replication. For this to happen, large and relatively stable populations are required and a mechanism which makes possible the transmission (and storage) of information, i.e., language, writing or symbolic thought. If you do not have such a mechanism, information is always in risk of getting lost. Not surprisingly then, in their authoritative work about the major transitions in evolution like the origin of eukaryotes, sex, multi-cellular organisms and social groups, John Maynard Smith and Eörs Szathmáry (1995) list the emergence of human (or hominid) language as the, for the time being, last big evolutionary transition, a breakthrough which, according to the authors, enabled the dispersal of *H. sapiens*, from Africa, throughout the world.

So for many smart animals like birds and primates it may not be the creative component which poses the real difficulty but rather the transmission component: they simply do not have a proper medium to transmit information efficiently. Thus the developmental psychologist Michael Tomasello (1999), co-director of Leipzig's Max Planck Institute for Evolutionary Anthropology, has aptly used the metaphor of a "ratchet" to explain why human (or hominid) culture is cumulative whereas animal culture is not. According to Tomasello, cumulative cultural evolution not only requires social learning and creative invention but also faithful social transmission that functions as a ratchet to prevent newly acquired information from "slipping back" (i.e., getting lost). The ratchet preserves information and ensures that modifications and improvements gradually accumulate over time. This picture suggests that such a ratchet (i.e., a proper transmission mechanism) creates some sort of threshold or hurdle which is difficult, if not impossible, to overcome by most animal species. As far as we know, we humans – and our hominid cousins and ancestors – are the only ones to have achieved this. We are the only true cultural animals.

In his seminal work on the cognitive aspects of human evolution, the psychologist Merlin Donald (1991) distinguishes three major stages in the origins of culture and the evolution of the human (or hominid) mind. The first stage is what he calls the advent of "mimetic culture" which gave our early ancestors the ability to communicate through manual or facial gestures – mime – and other voluntary motor acts. This communication system enabled them to share knowledge with learning taking place in the form of direct instruction and imitation. Mimetic culture was probably mastered two million years ago by *H. erectus*, our ancestor which accomplished the first exodus out of Africa. The second major step was the transition from mimetic culture to what Donald calls "mythic culture". In mythic culture spoken language plays a pivotal role because, as we have seen, language allows for a much more efficient mode of communication and information transfer. Donald argues that the telling of myths was the primary function of this prehistoric language, with the myths containing all kinds of accumulated knowledge about the world our ancestors lived in. Donald estimates that the ability to convey knowledge through mythic narrative is some 200,000 years old, coinciding with the earliest appearance of *H. sapiens* in Africa. Donald calls the third and, for the time being, last stage in the evolution of the hominid mind "theoretic culture" because it allowed our ancestors for the first time to reflect on their own knowledge. Theoretic culture is characterized by the fact that accumulated information was no longer restricted to the boundaries of the human brain, but could now be gathered in external storage systems, i.e., be written down in the form of cuneiforms, hieroglyphs or other symbols. Theoretic culture probably originated some 10,000 years ago in the Middle-East. It enabled our Neolithic ancestors to critically examine their own culture, thus establishing the first rudimentary forms of philosophy and science. Donald argues that this third transition has led to the greatest reconfiguration of cognitive structure in hominid history, but without much genetic change. Apparently genes do not make the difference, but accumulated culture does.

## 5. CULTURAL INHERITANCE AND FAMILY TREES

We have seen that for culture to evolve, a high-fidelity transmission mechanism is needed which carefully sieves, stores and accumulates information. Such a "ratchet" creates a process in which information gradually builds up to such an extent that the output or the products of this process can no longer be achieved by individual innovators. The products of cultural evolution, whether stone tools or high-end computers, typically exhibit intricate design which is the result of many generations of collaborative innovators standing on top of each other's shoulders. In other words, is seems that in cultural evolution each generation somehow "inherits" the achievements of previous generations. Several authors have argued that this process is intrinsically Darwinian in character since cultural evolution seems to be propelled by a similar mechanism as which underlies biological evolution, i.e., a process of variation, selection and transmission (Richerson and Boyd 2005; Mesoudi 2011). Combined these three ingredients create a cumulative selection process in which the outputs of each sieving stage serve as inputs for the next sieving stage, and so on. And as we all know, cumulative selection is the hallmark and underlying mechanism of Darwinian evolution.

The philosopher Daniel Dennett (1995) has argued that Darwinian evolution is "substrate neutral", i.e., the process is neutral with respect to the medium of evolution and neutral with respect to the entities that evolve. So in principle any dynamic "system" could evolve in a Darwinian manner as long as the three aforementioned ingredients are present: variation, selection, and transmission. In the case of human culture, this condition seems to be met. In the biological realm we find *genetic* variants which are generated, selectively retained and then transmitted to the next generation; in the cultural realm we find *cultural* variants (e.g., ideas and skills) which are generated, selectively retained and then transmitted to the next generation. So biological evolution is concerned with the differential distribution and changing frequencies of genetic variants, whereas cultural evolution is concerned with the differential distribution and changing frequencies of cultural variants. Yet one could object that the supposed analogy is misleading because biological evolution is unintentional and blind whereas cultural evolution seems, at least to a large extent, intentional and goal-directed. Therefore one should label cultural evolution as "Lamarckian" rather than "Darwinian". But plausible as this objection may seem, I believe that it is actually ill-founded because cultural evolution is "Lamarckian" only in a caricaturized sense. I will delve deeper into this matter later on.

Another obvious difference between the two kinds of evolution is that in the biological realm genetic information is usually transmitted through sexual and asexual reproduction, whereas in the cultural realm cultural information is transmitted through imitation, social learning and language. Yet in both cases the products of these processes, whether biological adaptations or cultural artifacts, typically bear the mark of a cumulative selection process: they exhibit complex and often functional design. Moreover, in both kinds of evolution we see the formation of recognizable lineages through time and the occasional splitting of one lineage into two or more branches, resulting in tree-like phylogenies or "family trees". Similar to evolutionary biologists, researchers studying cultural evolution are now able to analyze and reconstruct the lineages of such diverse cultural traditions like the North-American prehistoric stone tool and projectile point industries (O'Brien, Darwent and Lyman 2001), Polynesian canoe fabrication (Rogers, Feldman and Ehrlich 2009), the Indo-European language family (Gray and Atkinson 2003) and Italian violin design (Chitwood 2014), amongst others. In *The Descent of Man*, Darwin already noticed these remarkable analogies between biological and cultural evolution. With regard to language Darwin (1871: 60) writes: "The formation of different languages and of distinct species, and the proofs that both have been developed through a gradual process, are curiously parallel." Thus nowadays linguists analyze the historical relations between languages by looking, among other things, for cognates – i.e., similar words shared across languages like *nacht*, *notte*, *noche*, *nuit*, etc. – which could indicate common ancestry, just like biologists study the historical relations between different species by looking at homologies (shared traits in different species) which indicate common origin.

Some authors, most notably Dawkins (1976) and Dennett (1995; 2002), have argued that the similarities go even deeper. They posit a strong analogy between genetic variants (i.e., genes) and cultural variants which Dawkins calls "memes". A meme is a unit of cultural information which survives and reproduces by leaping from brain to brain. Both genes and memes are therefore "replicators" which encourage their own reproduction. Also, like genes, memes are involved in a constant struggle for survival in which some memes will increase in frequency while others will perish. And finally, both genes and memes are carriers of information. Genes carry instructions for building proteins and (parts of) organisms; memes carry instructions for building (parts of) culture. Moreover, both genes and memes can carry *latent* information because neither genes nor memes have to be expressed in order for them to be transmitted. A recessive gene may lie dormant for many generations before it expresses itself again in a phenotype. Similarly, a particular meme, say a piece of music, can be transmitted through the ages in sheet music without actually being performed.

Yet in other aspects memes may not be like genes at all. For instance, several critics have argued that, unlike genes – which are heritable and functional bits of DNA – , memes are usually not that discrete or particle-like (Richerson and Boyd 2005). Surely some memes can be relatively simple and hence neatly circumscribed like Pythagoras' theorem or the song *Auld Lang Syne*, but more often memes are quite complex and fuzzy like, say, the concepts "Asian Cuisine" or "Judaism". Richerson and Boyd (2005: 60) therefore argue that analogies must not be stretched too far. They write: "A Darwinian account of culture does not imply that culture must be divisible into tiny, independent gene-like bits that are faithfully replicated." They believe that memes – or "cultural variants" as they prefer the more neutral term – are only loosely analogous to genes. Cultural evolution may not always involve high-fidelity replication, nor do cultural variants always consist of tiny bits of information. But nevertheless, one could still argue that cultural evolution is fundamentally Darwinian in its basic structure. For even if there is no real cultural equivalent of genes, in the evolution of culture one can discern many Darwinian features like: variation, competition, selection, inheritance, cumulative change, geographical distribution, convergent evolution, changes in function, homologies and analogies, lineages and tree-like phylogenies, etc.

## 6. VARIABLES IN CULTURAL EVOLUTION

Because cultural evolution is a Darwinian cumulative selection process, researchers are able to employ the same mathematical models and techniques that are used to understand, explain and predict the outcomes of biological evolution. Computer simulations may for instance indicate whether a particular cultural variant will increase in the population or will soon disappear altogether, or whether two or more cultural variants are able to coexist or not, etc. By slightly changing the variables and the initial micro-evolutionary conditions, researchers can analyze the long-term macro-evolutionary consequences of their models. For example, the transmission of cultural information may depend on several modes of cultural inheritance. Perhaps the oldest and most common trajectory in which information is passed on or inherited is "vertical" or parent-to-offspring transmission. Other trajectories are "horizontal" transmission among peers form the same generation, and "oblique" transmission from members of one generation to members of a later generation as in the case of formal teaching. Mathematical models demonstrate that if vertical, downward transmission would be the main direction in which cultural information would flow, cultures remain rather static (Cavalli-Sforza and Feldman 1981). Hence for cultures to stay innovative and flexible, horizontal transmission is needed because new ideas should not only spread from parents to offspring, but to contemporaries as well.

Apart from different trajectories, there may also be differences in speed by which cultural information gets transmitted. Cavalli-Sforza and Feldman (1981) constructed models in which they analyzed the speed of evolution (i.e., the rate at which a new cultural variant spreads through the population) by manipulating the modes of transmission. Thus, as one would expect, when transmission is one-to-one or one-to-few, cultural evolution is relatively slow, whereas evolution gets faster when transmission becomes one-to-many. In small prehistoric populations of hunter-gatherers the mode of transmission was probably mainly vertical and one-to-few, resulting in a relatively slow pace of evolution. By contrast, in our modern world with its internet and social media the mode of transmission has dramatically shifted to horizontal and one-to-many, resulting in a constant and massive flow of information crisscrossing our planet and a corresponding high pace of evolution.

Finally, a third variable in cultural evolution is the potential bias by which information gets transmitted. For not all cultural variants are equal. For instance, a particular cultural variant may increase in frequency in a population, at expense of others, because of its intrinsic attractiveness. Technically this is called "content bias", i.e., the content of a variant affecting its probability of being transmitted. But there are many other ways in which transmission can be skewed. Let me briefly address two other types of bias here. A second type of transmission bias is called "frequency-dependent bias". This involves using the frequency of a trait as a guide as to whether to adopt it, irrespective of its content. A frequency-dependent bias will lead to conformity when people adopt the *most* common variant in the population, whereas it leads to nonconformity when people adopt the *least* common variant in the population. Conformity – when in Rome, do as the Romans do – offers an explanation for long-lasting cultural traditions, even when migration rates are high (Boyd and Richerson 1985). Yet another type of transmission bias is called "prestige bias" where the identity, and fame, of the person from whom the cultural variant is acquired affects the variant's success. Prestige bias might not be a bad strategy because by imitating the behavior of successful individuals in the population, you have a chance to become successful yourself (Henrich and Gil-White 2001). Not surprisingly then, our propensity to admire and imitate famous people, like rock idols and movie stars, is exploited by the world of advertising by presenting celebrities as role models. If a commercial shows, say, Claudia Schiffer driving a certain brand of car you might get tempted to buy one too. As was mentioned above, by manipulating the aforementioned and other variables in models, researchers are now able to simulate the patterns and outcomes of cultural evolution. In fact, it is worth pointing out that the mathematical models and techniques currently being used are largely identical to those being employed during the 1920s and 1930s during the "Modern Synthesis" when scientists like R.A Fisher, J.B.S. Haldane and Sewall Wright lay the foundations for population genetics, one of the cornerstones of modern evolutionary biology.

But apart from being a purely academic enterprise, the study of cultural evolution may soon draw attention in the business world as well – and for good reason. Our digital society generates an enormous and increasing amount of data which already caught the eye of internet giants like Facebook and Amazon and "big data" companies like Google, IBM and Oracle because they realize that the raw data contain real treasures of information. For instance, by analyzing the daily flow of 500 million tweets, big data companies are now able to predict patterns and to spot hypes and trends, sometimes even before the participants themselves have noticed them (Choi and Varian 2012). In a similar vein, by analyzing big data we may predict a crash of the stock market, a sudden change in public opinion, or the outbreak of a political revolution even before the people involved are aware of it (such feats of forecasting the present are also known as "nowcasting", a term borrowed from meteorology). Also remarkable is that within the current research on big data, such state-of-the-art tools as evolutionary algorithms and genetic programming are being employed in order to manage the flow of information and to detect meaningful and predictive patterns in it (Freitas 2002). Evolutionary computation may solve the problem of how to harness big data. Some researchers believe that the rapidly evolving world of big data analytics, machine learning and data-driven science will eventually transform the way we work, live and think (Lohr 2015). If that is true, the research on big data may further advance and complement the science of cultural evolution as well.

## 7. IS CULTURAL EVOLUTION REALLY DARWINIAN?

In this final section we must address the issue already briefly mentioned above, that is, the question whether cultural evolution really qualifies as "Darwinian" or not. For one could easily object that because cultural evolution is – at least to a certain extent – directional and because most cultural variants are – at least to some degree – purposively generated and transmitted, we should call the process "Lamarckian" rather than "Darwinian". In other words, cultural evolution must be Lamarckian because Lamarckian evolution is "guided" whereas Darwinian evolution is "blind" (Kronfeldner 2007; Mesoudi 2008). As is well-known, Lamarckian evolution is guided or directed because new variants are already adjusted to the "needs" they have to fulfill. Because in Lamarckian evolution acquired characters are inherited, evolution is able to anticipate the challenges posed by the environment, thus rendering Lamarckian evolution swift and progressive. Hence in Lamarckian evolution a wasteful and cumbersome selection process which retains the occasional useful variants is not needed because the variants are already pre-adapted to their upcoming task. In Darwinian evolution, by contrast, such a selection process is an essential feature because in Darwinian evolution new variants cannot anticipate the challenges posed by the environment, i.e., they are blind.

Now I suppose that nobody would claim that cultural evolution is "Lamarckian" in a literal sense, for that would mean that the ideas and skills that we acquire during our lifetime should somehow become incorporated and encoded into our DNA and then be genetically transmitted to our offspring. That is probably not what people have in mind when they say that cultural evolution is "Lamarckian". Obviously they believe that cultural evolution is Lamarckian only in a *metaphorical* sense. That is, they simply mean that ideas and skills acquired during our lifetime can be transmitted to our offspring, through teaching, etc. But if that is what it takes to qualify as "Lamarckian", the statement becomes rather trivial. With the biologist and philosopher David Hull (1988) one suspects that people who claim that cultural evolution is Lamarckian actually have a caricaturized sense of the term in mind – they want to emphasize that cultural evolution, unlike biological evolution, is intentional. Now, of course, nobody would deny that cultural evolution indeed *is* to a large extent intentional, but the question then becomes whether intentionality can circumvent a wasteful Darwinian selection process. I think it cannot. Obviously cultural evolution generally involves the participation of conscious and intelligent agents, but the fact that much of culture is consciously and deliberately generated does not guarantee that, for that reason, it will be selectively retained. Despite its apparent goal-directedness, cultural evolution may ultimately be as wasteful and laborious as biological evolution. Surely cultural (and scientific) evolution can be made to *look* very directional and efficacious by careful editing, since it is only in retrospect – when all the failures and dead ends are rubbed out – that we get the impression that cultural evolution is a smoothly guided process. In reality, though, we will always need the hindsight of a selective system to separate the wheat from the chaff (Buskes 2013). In short, the fact that a particular cultural variant is generated intentionally is not a sufficient condition for that variant's success, and it might not be a necessary condition either because some cultural variants, like certain customs or pronunciations, are *not* deliberately generated – they just crop up – but nevertheless may turn out to be highly successful. So even if we grant that cultural evolution is intentional and driven by intelligent agents, it is not clear what we would gain by calling such a process "Lamarckian".

In the end this issue perhaps boils down to a matter of definition. What is the defining characteristic or key signature of Darwinian evolution? Is it blind variation or is it cumulative selection? I believe it is the latter. Let me explain. Suppose that, for some reason, new variants in Darwinian biological evolution would suddenly be guided instead of blind. Would that fundamentally change the overall process of evolution in any way? It probably would not because guided variation would only *increase* the efficiency of such a process. So guided variation is still variation in need for subsequent selection. It is well-known that Darwin drew an analogy between artificial selection and natural selection. Darwin discovered the principle of natural selection by looking at how farmers and breeders have improved their stock and crops by selecting individual animals and plants with certain desirable characteristics. Thus by selecting particular individuals for further propagation, farmers and breeders restricted the range of variation, i.e., they made the variation guided. Yet despite the fact that such variation is guided and that selective breeding is utterly intentional and goal-directed, Darwin treated artificial selection as a "special case" of natural selection, seeing no fundamental difference between the two processes. Both artificial and natural selection are cumulative sieving processes and it is this feature that makes artificial selection, despite all intentions and manipulations, a proper instance of Darwinian evolution.

Nevertheless, in the contemporary literature on cultural evolution, a term like "Lamarckian inheritance" is still being used, albeit not in the literal biological sense but rather in a metaphorical manner. In the terminology of Boyd and Richerson (1985) "Lamarckian inheritance" refers to a mode of transmission in which one individual acquires information form a second individual, modifies that information in accordance with his or her own views, and then transmits that modified information to other individuals in the population. Yet despite these Lamarckian labels, Boyd and Richerson, as well as many other researchers involved in the science of cultural evolution, stress that the overall process of cultural evolution must be viewed as Darwinian. The basic assumption is that all instances of complex cultural design are the result of cumulative sieving processes. Not foresighted variation but hindsighted selection is the secret to creative Darwinian evolution.

## 8. CONCLUSIONS

In this article I have sketched the early origins, the recent growth and the theoretical contours of the emerging science of cultural evolution. Although the first initiatives on behalf of this project can already be found in Darwin's work, the science of cultural evolution only began to gain momentum during the second half of the 20$^{th}$ century when several taboos and prejudices had been cleared away. During the last few decades a new, multidisciplinary research program has emerged that concerns itself with the evolution of culture. The basic idea is that cultural evolution, like biological evolution, is essentially a Darwinian cumulative winnowing process in which some cultural variants are selectively retained and information gradually builds up. Yet although culture is quite common, even among animals, cultural evolution is remarkably rare. Indeed, as far as we know, we are the only species which developed a cumulative, evolving culture. Why that is the case is still a bit of a mystery. Perhaps the key to understanding this riddle is to focus on the particular transmission mechanisms which propel human cultural evolution, most notably our language. Our language created a new medium or channel through which information could be faithfully transmitted to next generations without getting lost. Animals lack this typical "ratchet effect" of human cultural evolution. In human cultural evolution the transmission or "inheritance" of information also results in tree-like phylogenies so characteristic of Darwinian evolution. Researchers are now able to reconstruct such cultural lineages in the same way as evolutionary biologists try to reconstruct the tree of life. By recognizing the fact that human culture has its own inheritance system, which can influence and even alter the path of genetic-biological evolution, researchers now have the proper theoretical tools to understand the impact and long-term consequences of culture. The science of cultural evolution may be scarcely out of the egg, but the prospects are promising. The research program clearly exhibits theoretical and empirical progress which, according to the philosopher of science Imre Lakatos (1977), can be taken as a sign of a scientific discipline in bloom. Perhaps we are witnessing the dawn of a new scientific era, an era in which the Darwinian paradigm will reach its pinnacle and full potential.

**Author**

**Chris Buskes** finished his BA and MA in philosophy at Radboud University, Nijmegen, The Netherlands. In 1998 he earned his PhD in Philosophy of Science at the same institution. At present he is Assistant Professor at the Department of Philosophy at Radboud University.

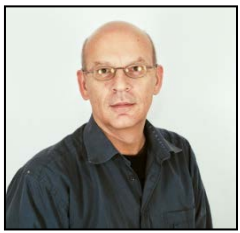